\documentclass{article}

\usepackage[preprint]{neurips_2022}

\usepackage[utf8]{inputenc} 
\usepackage[T1]{fontenc}    
\usepackage{hyperref}       
\usepackage{url}            
\usepackage{booktabs}       
\usepackage{amsfonts}       
\usepackage{nicefrac}       
\usepackage{microtype}      
\usepackage{xcolor}         
\usepackage{graphicx}
\usepackage{enumitem}
\usepackage{caption}
\usepackage{subcaption}
\usepackage{comment}

\title{Learn2Trust: A video and streamlit-based educational programme for AI-based medical image analysis targeted towards medical students}

\author{%
  Hanna Siebert, Marian Himstedt and Mattias Heinrich\thanks{This work was in part funded by the German Ministry of Research and Education under grant 16DHBQP052} \\
  Medical Deep Learning Group\\
  University of Lübeck\\
  Lübeck, Germany \\
  \texttt{heinrich@imi.uni-luebeck.de} \\
}

\begin{document}

\maketitle

\begin{abstract}
 In order to be able to use artificial intelligence (AI) in medicine without scepticism and to recognise and assess its growing potential, a basic understanding of this topic is necessary among current and future medical staff. Under the premise of "trust through understanding", we developed an innovative online course as a learning opportunity within the framework of the German \textit{KI Campus} (AI campus) project, which is a self-guided course that teaches the basics of AI for the analysis of medical image data. The main goal is to provide a learning environment for a sufficient understanding of AI in medical image analysis so that further interest in this topic is stimulated and inhibitions towards its use can be overcome by means of positive application experience. The focus was on medical applications and the fundamentals of machine learning. The online course was divided into consecutive lessons, which include theory in the form of explanatory videos, practical exercises in the form of Streamlit and practical exercises and/or quizzes to check learning progress. A survey among the participating medical students in the first run of the course was used to analyse our research hypotheses quantitatively.

\end{abstract}

\section{Introduction}
Research in artificial intelligence has opened up a number of important new use cases for automatic data analysis in medicine within just a few years. Due to the rapid development, textbooks for basic methods, e.g. in the field of machine learning, are quickly outdated and open-access online courses (especially Stanford University CS231n [1]) play an important role in knowledge transfer.
Artificial intelligence will play an increasingly important role in the clinical practice of physicians. The use of AI in medical image analysis can lead to increased efficiency, e.g. through increased safety and speed, of diagnosis and therapy decisions. Even medical professionals without prior AI expertise can successfully understand and use AI with little learning effort [2].
However, the practical use of software based on artificial intelligence is often still fraught with mistrust for physicians and medical students. In contrast to conventional computer support, it is difficult for users to understand how a decision is reached. The decision-making process of AI algorithms is usually not interpretable for the human user, which has led to the designation of AI algorithms as "black boxes". This triggers reservations about the use of AI in medicine among many medical professionals, which should be counteracted through better understanding in order to be able to use the potential of these algorithms to improve medical care [3].
Therefore, our goal is to make the basic methods for training neural networks with the specific challenges of medical image interpretation learnable, understandable and thus more trustworthy for a large group of interested people.

\section{Methods and Materials}
Within the scope of the project, new didactic concepts for teaching complex technical-informatics content of artificial intelligence (AI) to interested medical students were successfully developed. By combining learning videos with interactive visual application examples with the help of a browser-based user interface (UI) for the underlying python program code (Streamlit), a completely new course concept could be established within the framework of an elective course of the medicine degree at the University of Lübeck. The learning offer comprised 5 modules, three of which were implemented in interactive Streamlit demos, which in turn were divided into up to 5 sub-chapters. The interactive topics included: 
\begin{enumerate}
    \item introduction, basics of AI
    \item (convolutional) neural networks
    \item medical image datasets
    \item classification
    \item semantic segmentation
\end{enumerate}

In two pilot studies, >150 students (spread over two semesters) were guided through the self-explanatory AI tasks in Streamlit either by online or face-to-face support. An empirical survey among the participants investigated the expectations, previous experiences and confidence that prospective doctors associate with the use of AI in research and clinical practice. It was found that better \textbf{Learning} leads to more \textbf{Trust} in the responsible use of AI and data in the clinic, also with regard to patients.
\subsection{Exemplary course material}
The following figures in Fig.~\ref{fig:material} demonstrate some examples from our Streamlit course material. We provide simple graphics to illustrate architecture, functionality, and evaluation of AI models for medical image analysis. Besides that, we use short explanatory texts, meaningful commented code snippets, and visualisations of medical application examples. Interactivity is achieved through checkboxes, quizzes, and sliders that allow it to visualise with multiple settings.

\begin{figure}[tbh]
  \centering
  \includegraphics[width=0.98\textwidth]{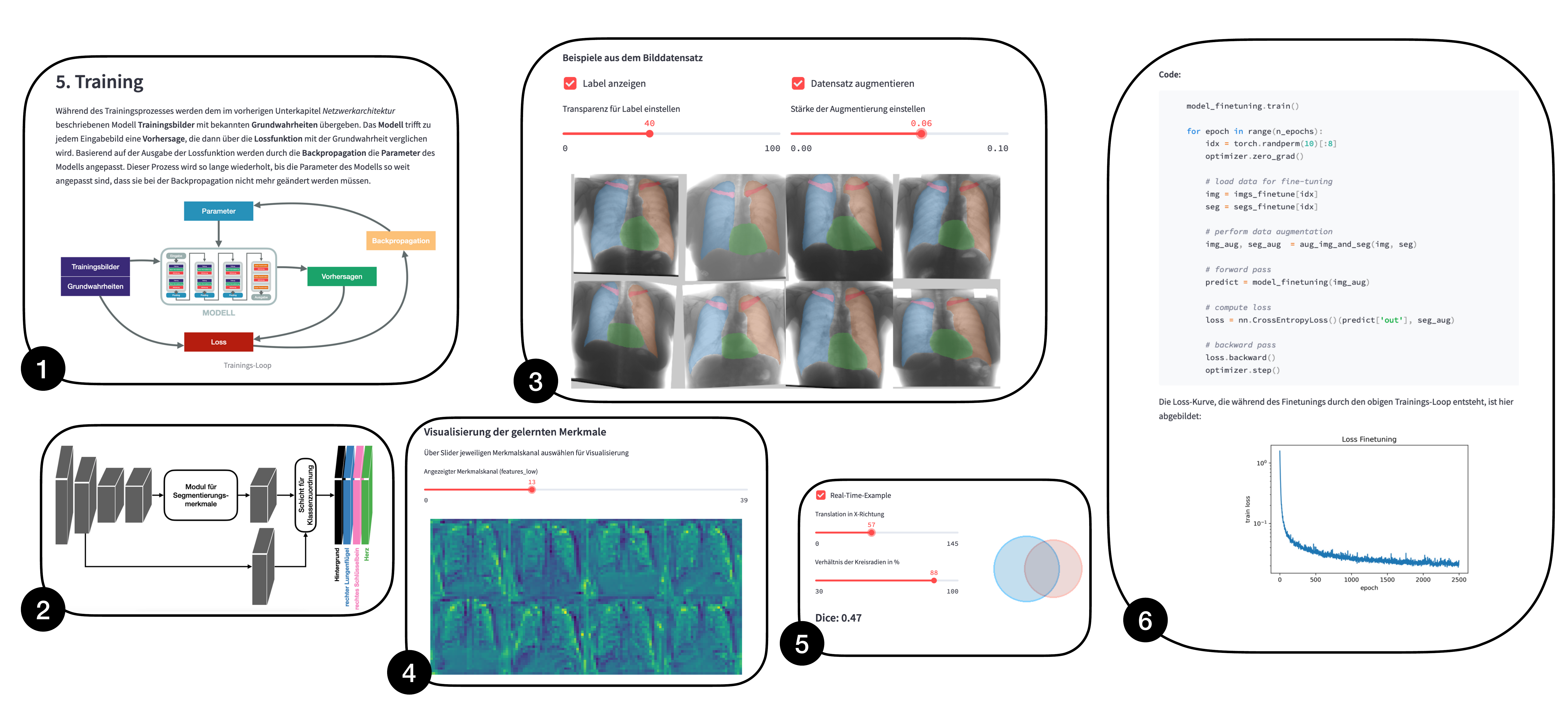}
  \caption{Expamples from our Streamlit course material: We provide explanatory texts (1), simple graphics (1,2), interactivity through siliders and checkboxes (3-5), and code snippets (6).}
  \label{fig:material}
\end{figure}

\subsection{Research hypotheses}\label{subsection:research_hypotheses}
Within the project, we addressed a number of research questions to investigate how successful our learning offer is in increasing the understanding of AI in medical image analysis and to what extent gains in understanding are associated with gains in confidence and willingness to use AI in medicine.
\begin{itemize}
    \item How well can basic understanding of AI in medical image analysis be increased among non-AI experts?
\item Does better understanding increase trust in AI in medicine?
\item Does better understanding increase the willingness to use AI in medicine?
\item Do learners feel better able to recognise possible sources of error in the use of AI in medical image analysis after taking part in the course? 
\end{itemize}
With these hypotheses in mind, we investigated how much the subjectively perceived basic understanding of AI in medical image analysis can be increased by participating in our learning offer. Furthermore, we wanted to determine how and whether learners' confidence in AI in medicine can be increased through our learning offer and to what extent gains in understanding are associated with gains in confidence. Our aim was also to study whether medical professionals are more willing to use AI in their daily work due to a better understanding of AI in medical image analysis and the increased confidence that potentially comes with it. Since scepticism about the use of AI in medicine is often caused by ignorance of possible sources of error, we finally investigated to what extent our learning offer can contribute to learners feeling able to recognise and assess sources of error.

\section{Questionnaires and Results}
\label{sec:questionnaires}
During our first run of the course, we issued questionnaires to the participants which included ten questions to analyse our research hypothesis. We received feedback from $33$ participants who indicated that they completed an average of $88,94~\%$ of the provided course materials.

The questions were as follows:
\begin{itemize}
\item Three questions about interest and prior knowledge:
\begin{itemize}[leftmargin=1cm]
    \item [(Q1)] I like to look more closely at technical systems.
    \item [(Q2)] I am interested in the topic of "Artificial Intelligence in Medicine".
    \item [(Q3)] I already had prior knowledge about how AI models work.
\end{itemize}
\item Three questions about personal attitude to the topic:
\begin{itemize}[leftmargin=1cm]
    \item [(Q4)] Trust in AI can increase the willingness to use it in medical image analysis.
    \item [(Q5)] Learning about AI in medicine can increase trust in AI.
    \item [(Q6)] I am willing to use AI methods of medical image analysis in my (future) professional life.
\end{itemize}
\item Four questions about the course "Learn2Trust":
\begin{itemize}[leftmargin=1cm]
    \item [(Q7)] The course "Learn2Trust" has increased my understanding of AI in medicine.
    \item [(Q8)] Through the course "Learn2Trust" I have become less sceptical about the use of AI in medicine.
    \item [(Q9)] Through the course "Learn2Trust" I feel able to recognise and assess sources of error when using AI in medical image analysis.
    \item [(Q10)] The contents of the course "Learn2Trust" have been clearly underchallenging/ slightly underchallenging/ optimally challenging/ slightly too challenging/ clearly too challenging.
\end{itemize}
\end{itemize}

For (Q1)-(Q9) we offered the answer options "totally agree", "agree", "little agree", "little disagree", and "totally disagree". Furthermore, the option "no answer" could be selected for each question.

\begin{figure}[tbh]
    \centering
    \begin{tabular}{ccc}
        \includegraphics[width=0.32\textwidth]{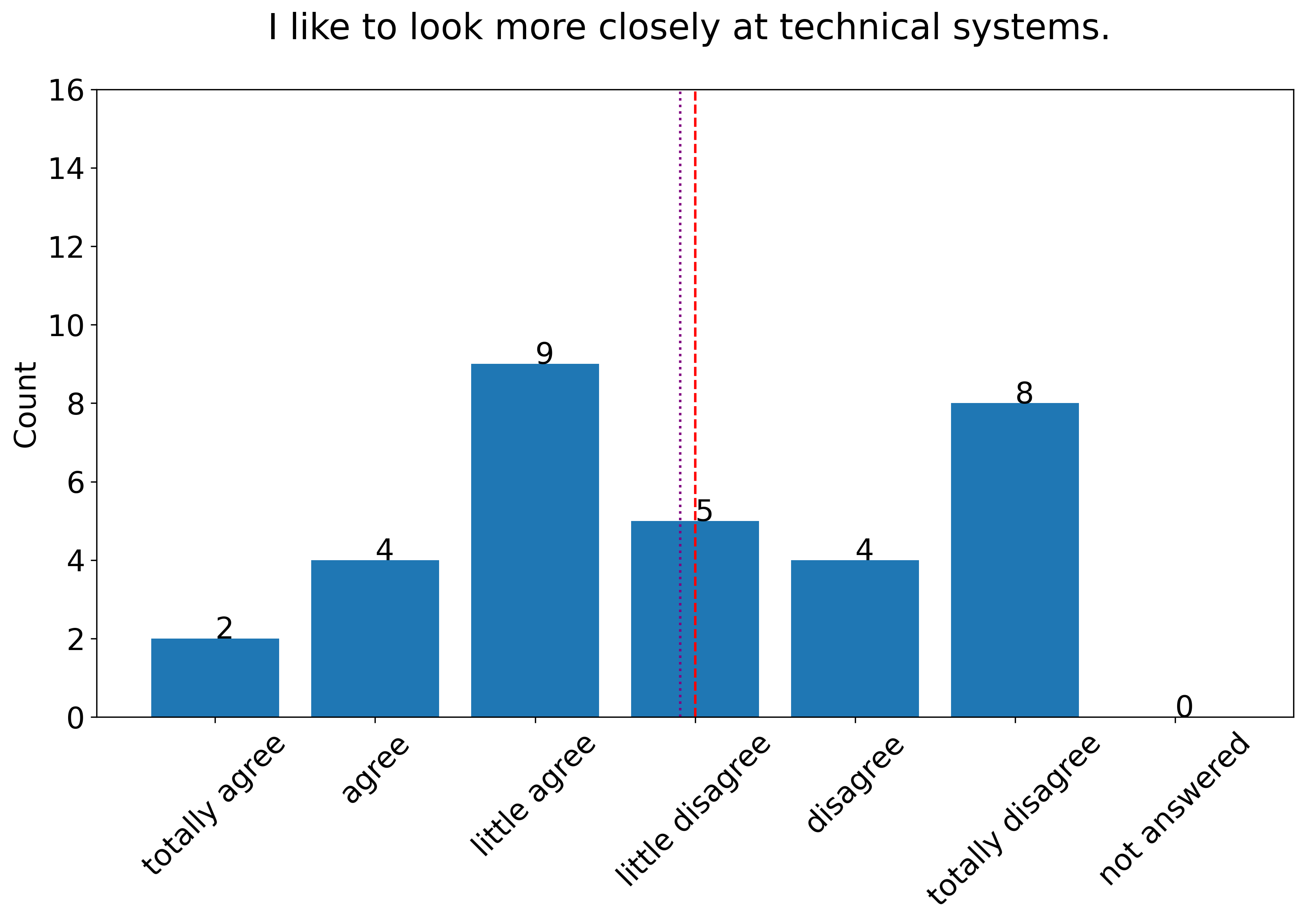} &  
        \includegraphics[width=0.32\textwidth]{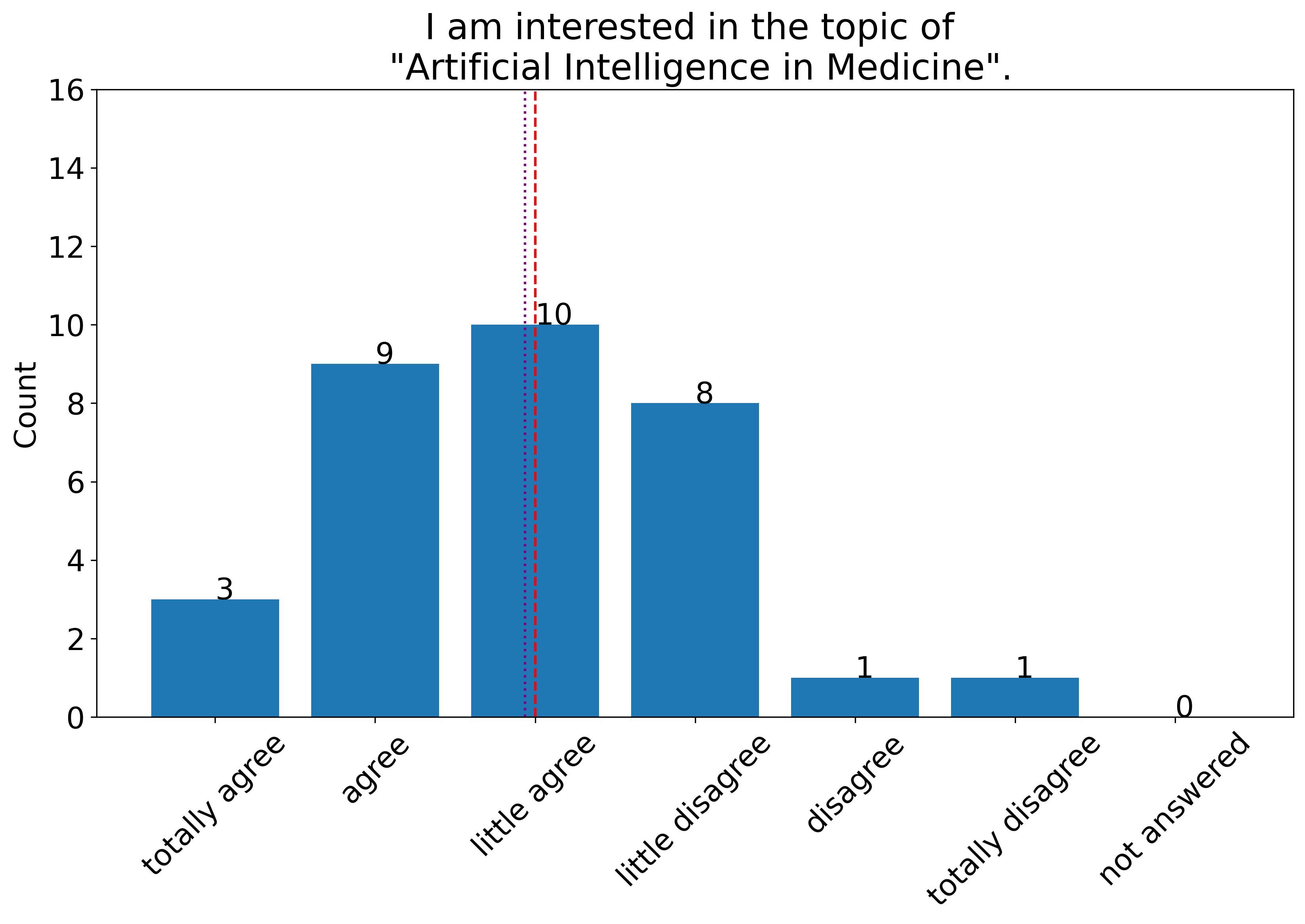}& 
        \includegraphics[width=0.32\textwidth]{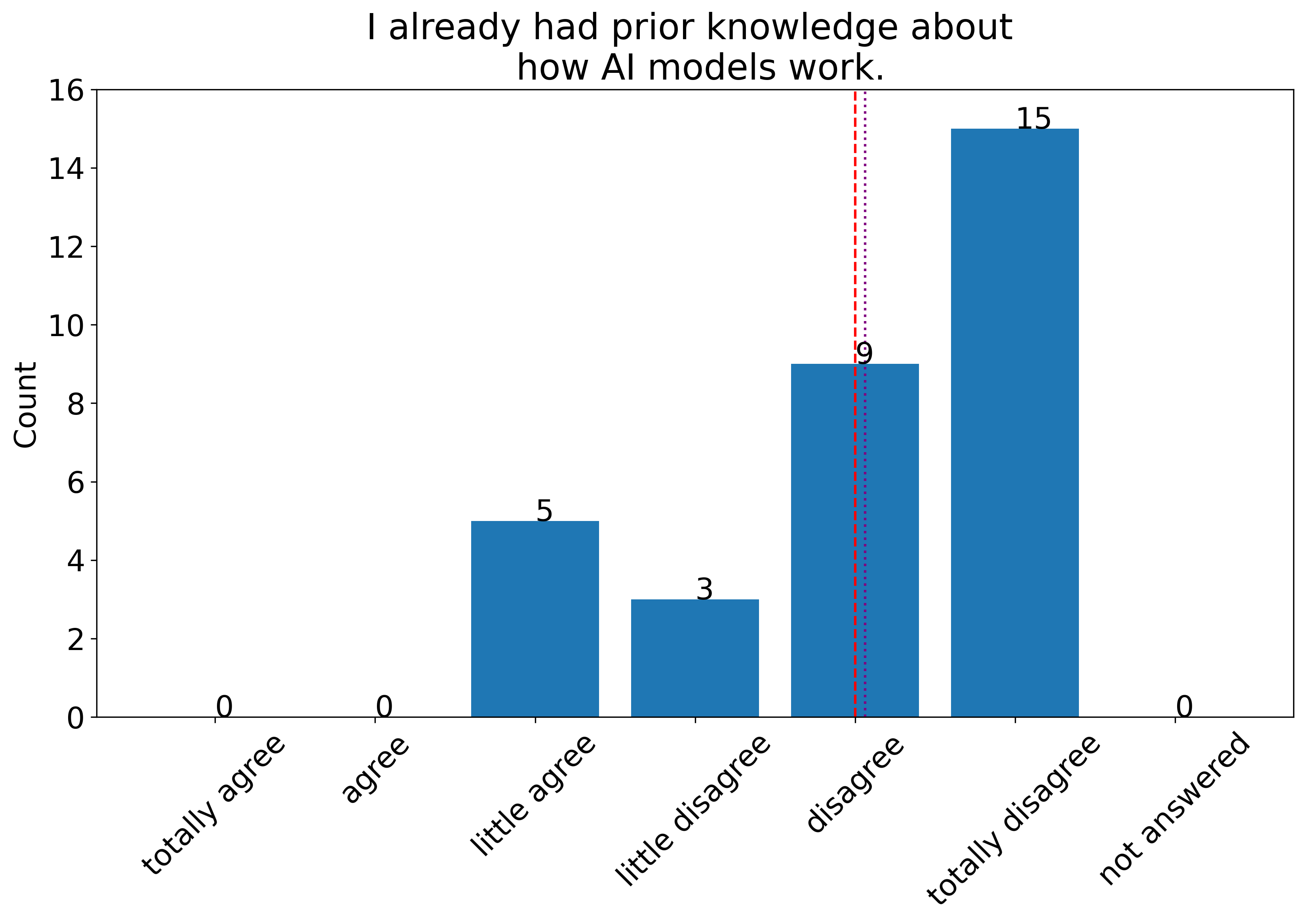}  \\
        (Q1) & (Q2) & (Q3)\\
        &&\\
        \includegraphics[width=0.32\textwidth]{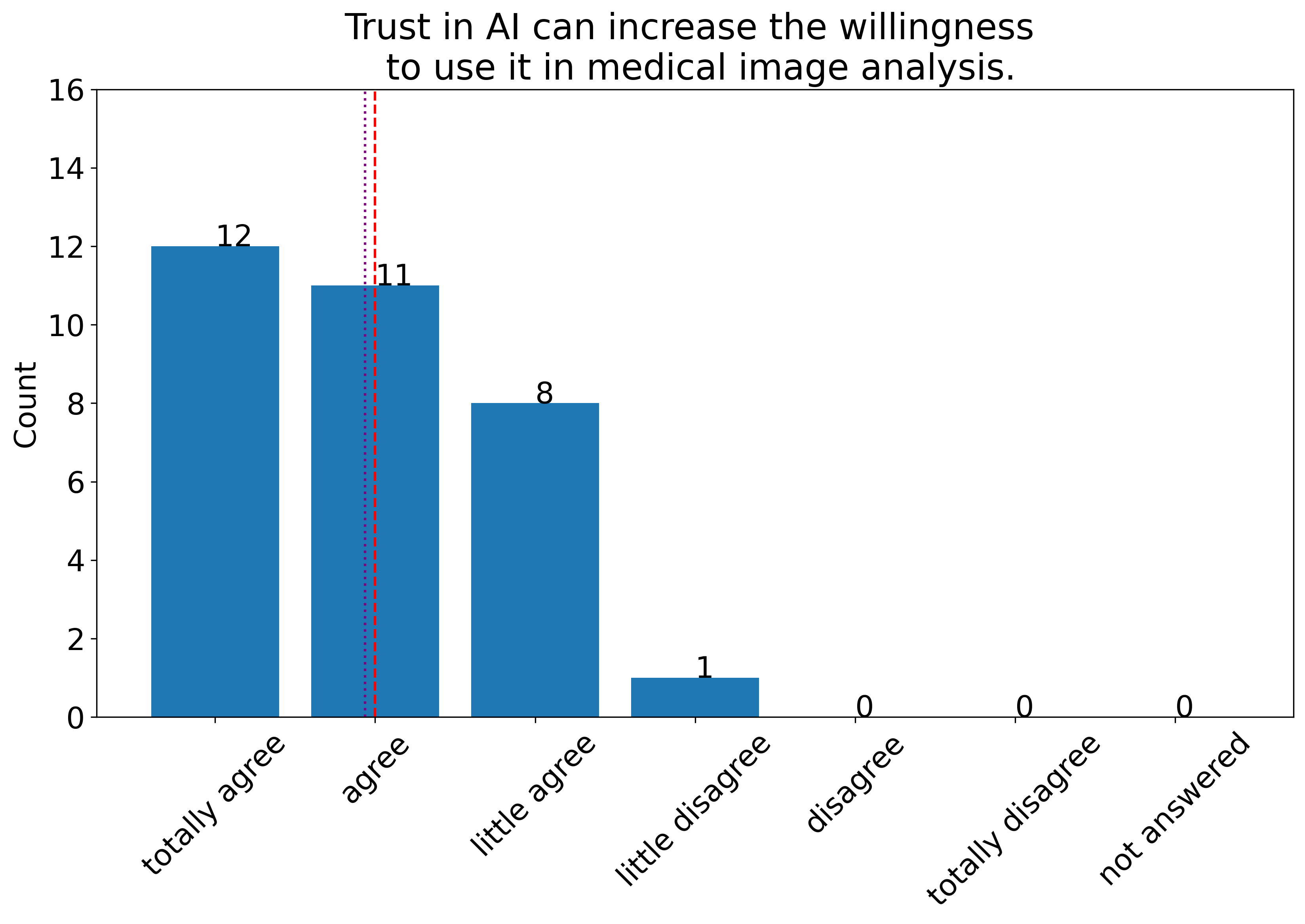} &  
        \includegraphics[width=0.32\textwidth]{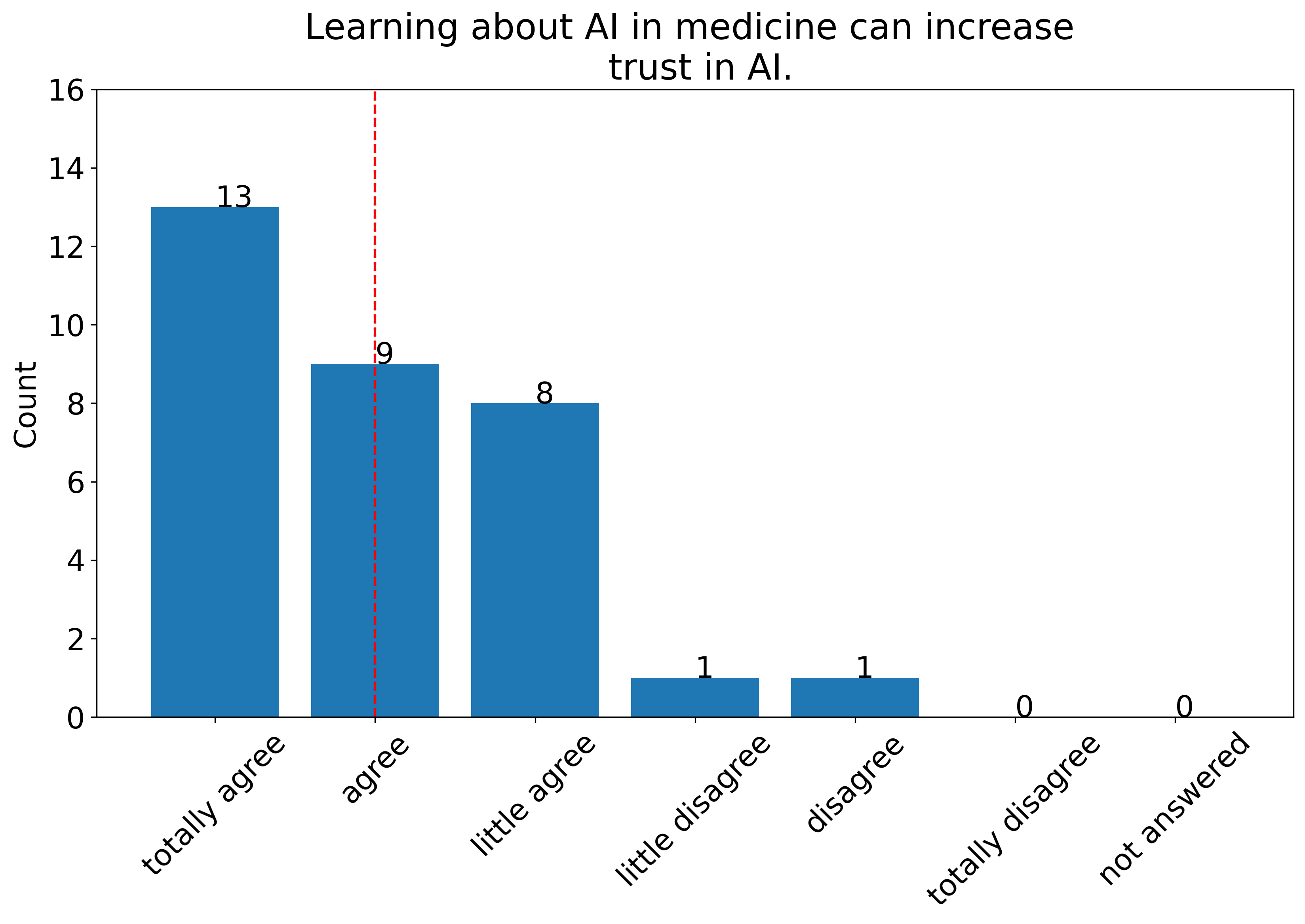}& 
        \includegraphics[width=0.32\textwidth]{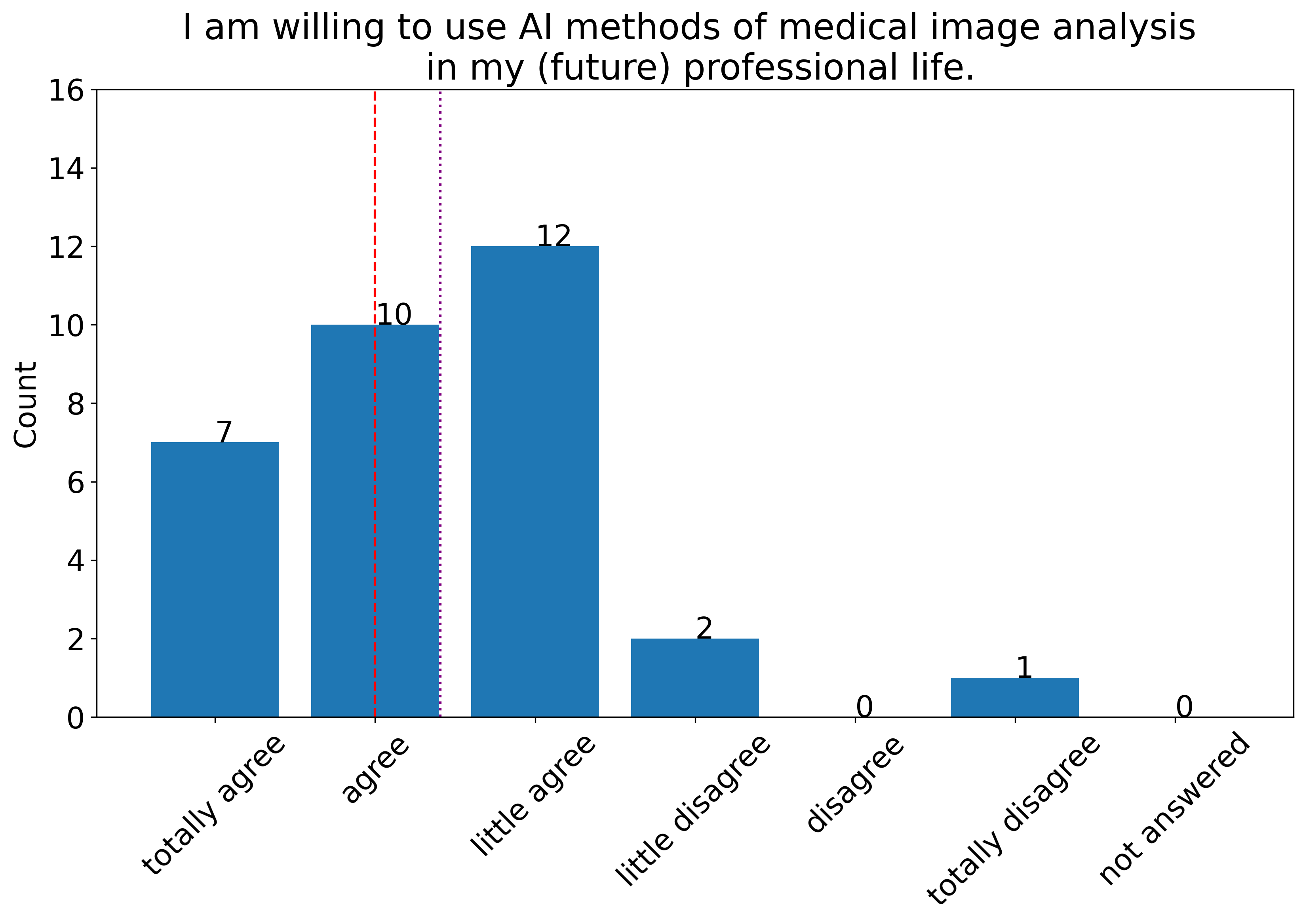}  \\
        (Q4) & (Q5) & (Q6)\\
        &&\\
        \includegraphics[width=0.32\textwidth]{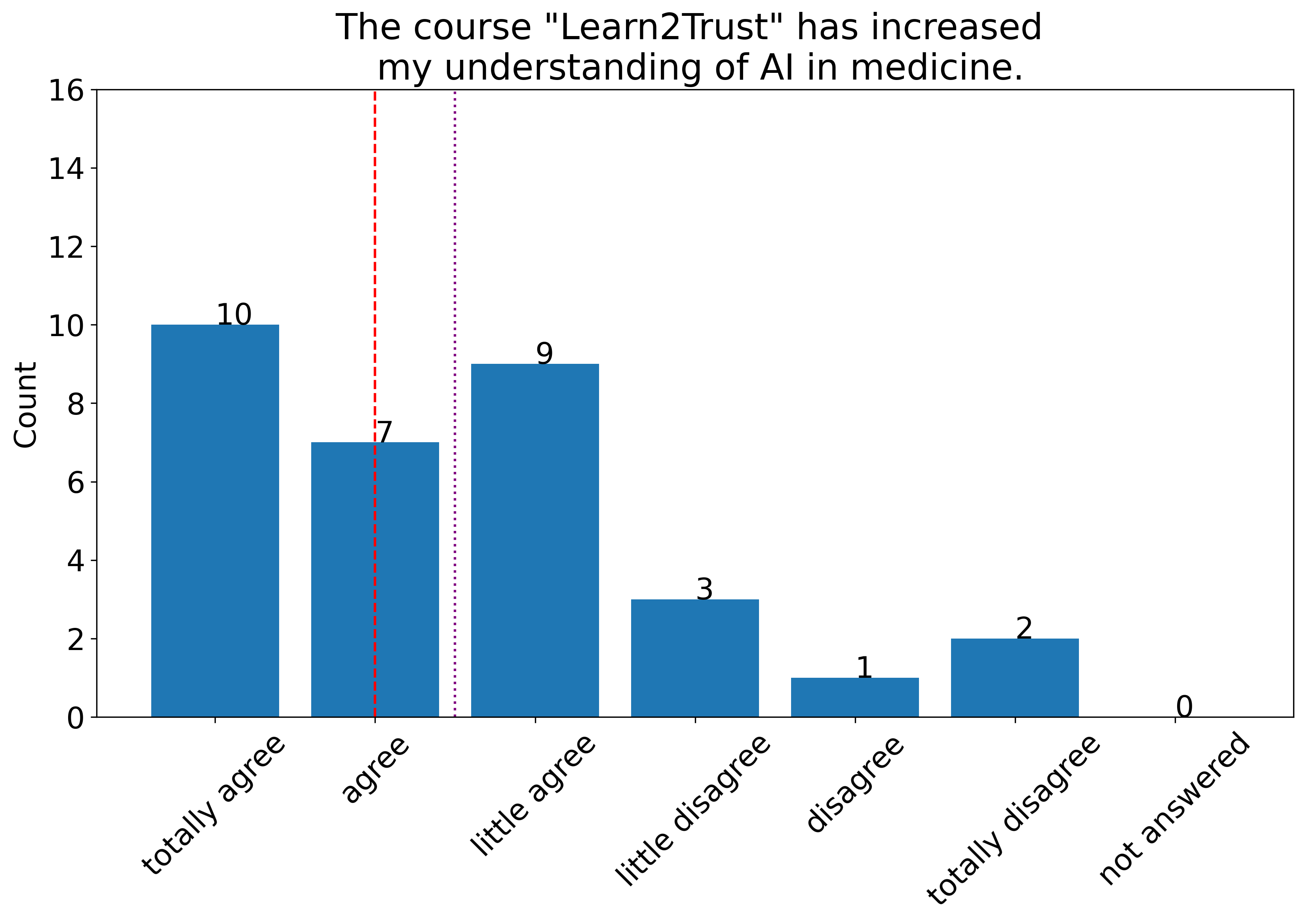} &  
        \includegraphics[width=0.32\textwidth]{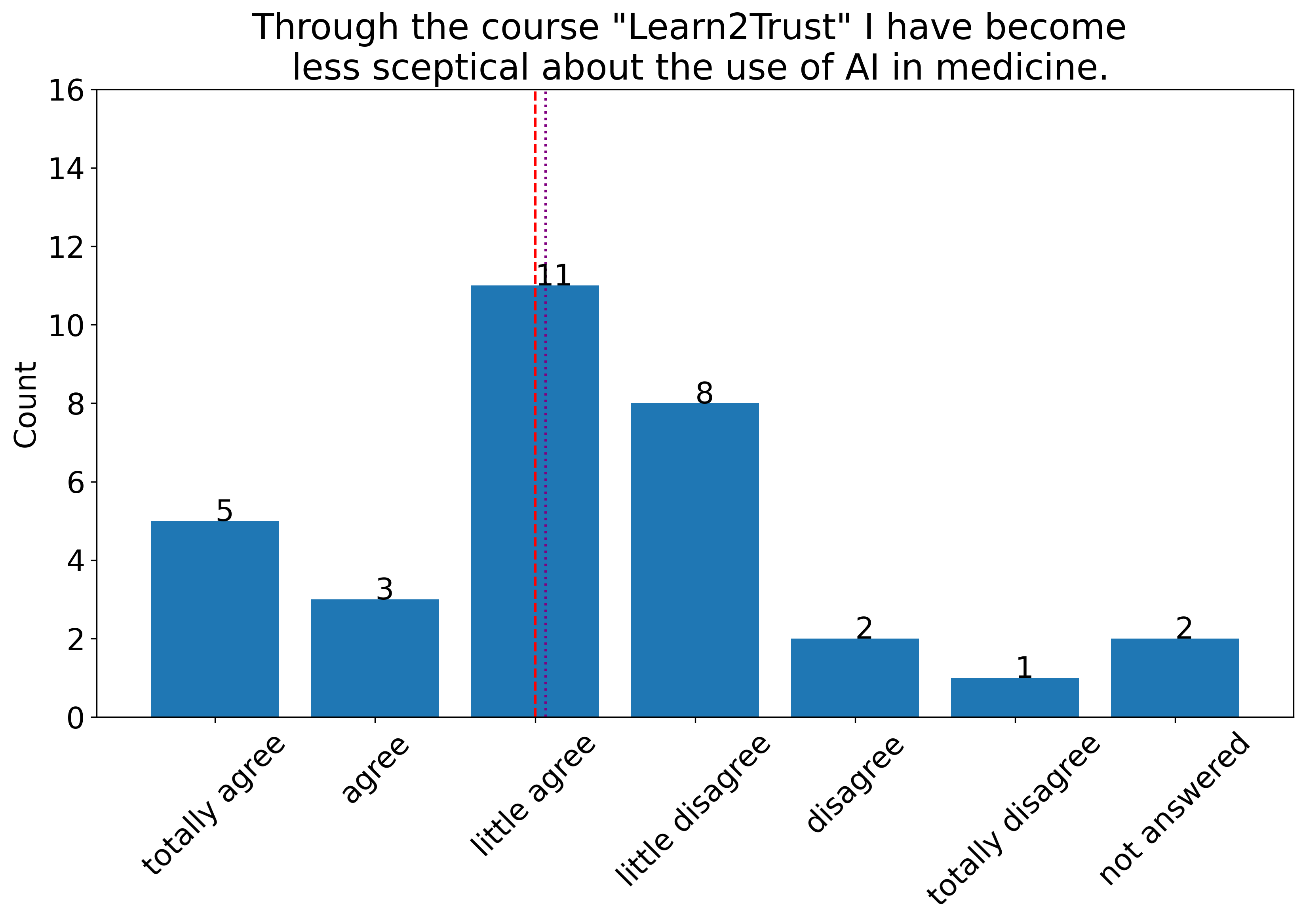}& 
        \includegraphics[width=0.32\textwidth]{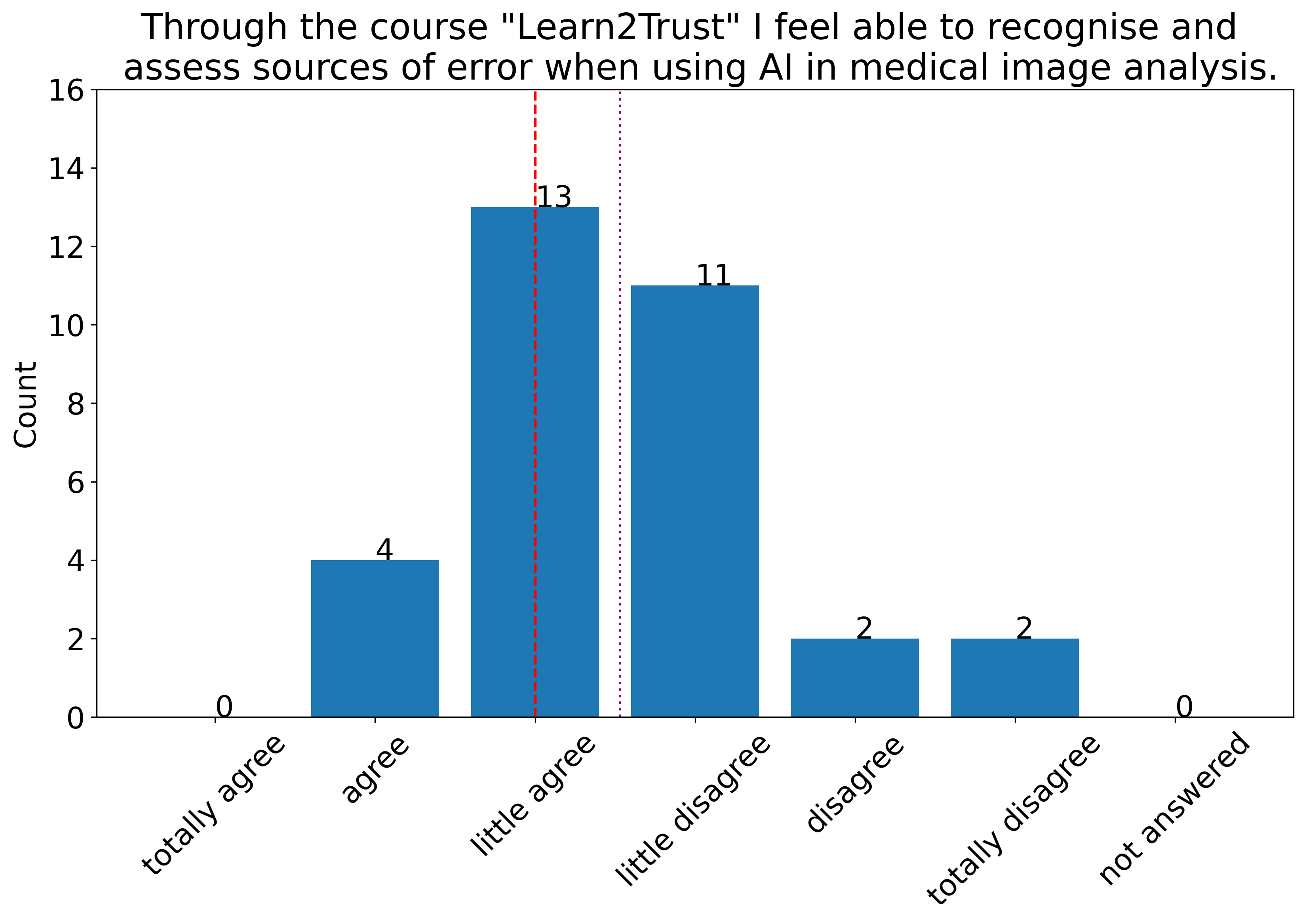}  \\
        (Q7) & (Q8) & (Q9)\\
        &&\\
        \includegraphics[width=0.32\textwidth]{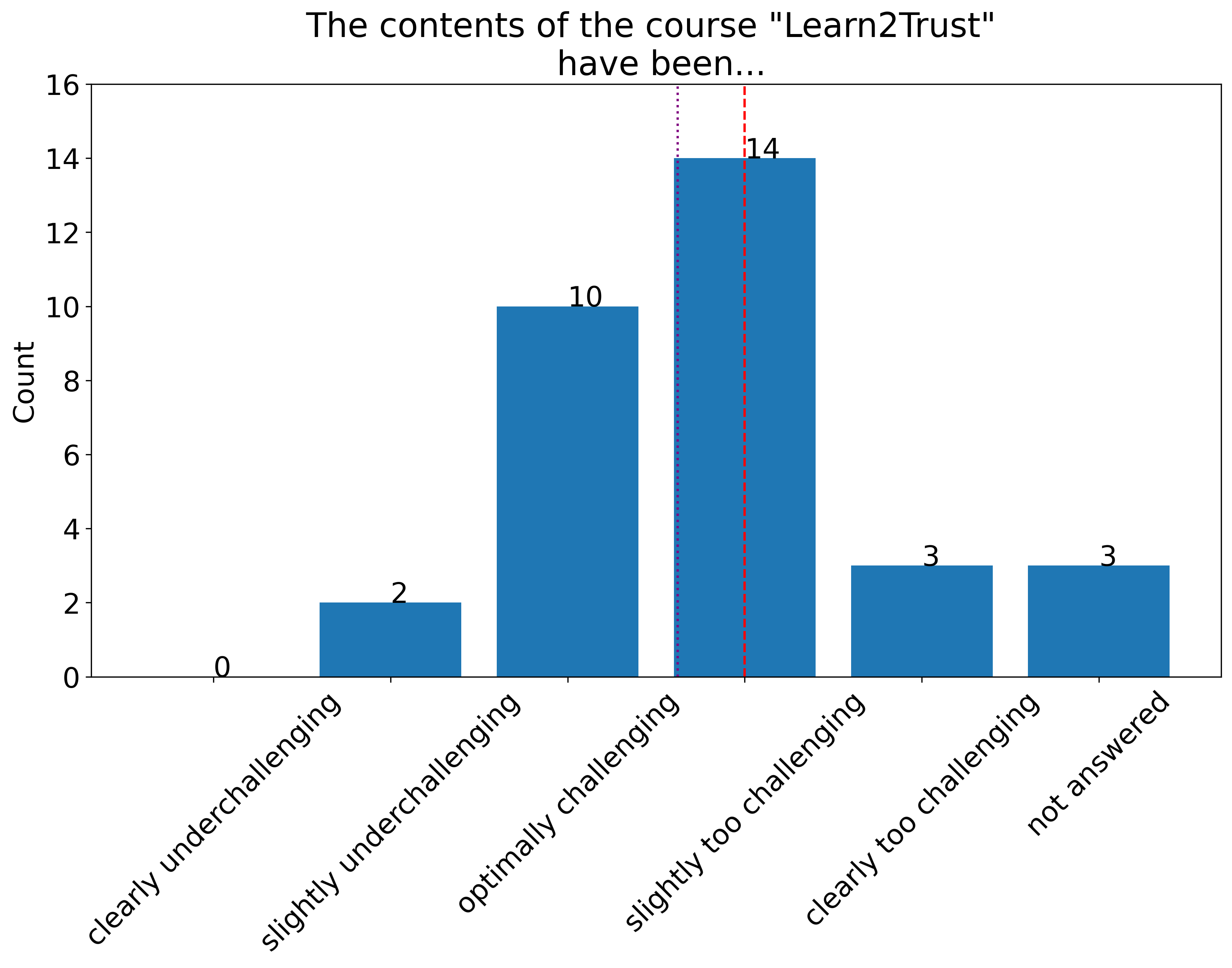} &
        \multicolumn{1}{l}{\includegraphics[height=0.23\textwidth]{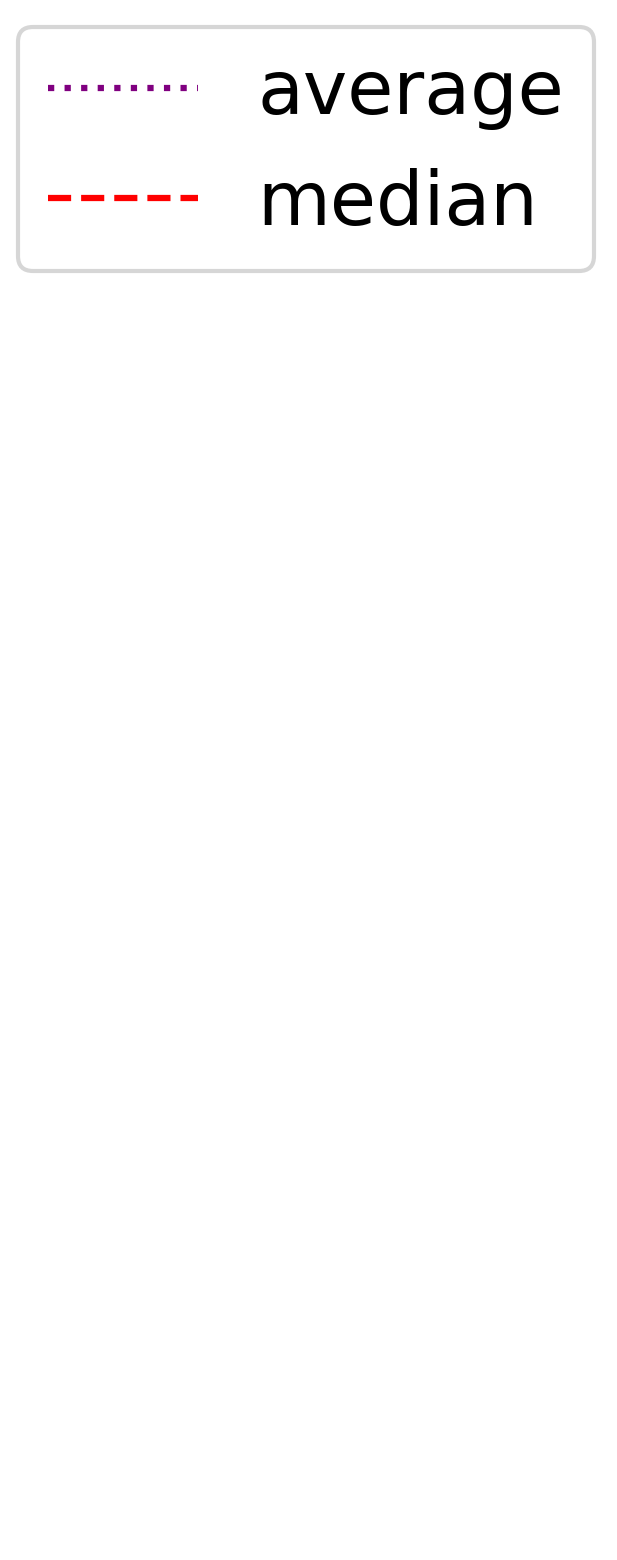}} & 
        \\
        (Q10) &  & 
    \end{tabular}
    \caption{Evaluation of the questionnaires: We indicate the number of given answers for each answer option for each question. Additionally, we give the average and median selected answer option (answer option "no answer" excluded; red: median; purple: average).}
    \label{fig:questionnaires}
\end{figure}

In Fig.~\ref{fig:questionnaires} we show the evaluation of our questionnaires. For each question and answer option, we indicate the number of given answers. We also give the average and median selected answer option (answer option "no answer" excluded).

The evaluation of the questions about interest and prior knowledge showed a trend towards interest in the topic of "Artificial Intelligence in Medicine" together with rather little existing affinity for technology and no prior knowledge about AI models. The evaluation of the questions regarding personal attitude towards AI in medicine, on the other hand, showed that the participants are willing to use AI methods for medical image analysis in their (future) professional life and a belief in the statements that learning about AI in medicine can increase trust in AI and that in turn trust in AI can increase the willingness to use it in medical image analysis. The feedback on the course "Learn2Trust" reported that the course has helped the participants to increase their understanding of AI in medicine. Regarding the questions whether the course has helped the participants to become less sceptical about the use of AI in medicine and whether the course has led the participants to feel able to assess sources of error when using AI in medical image analysis, the participant's most frequent answer has been "little agree". Moreover, the participants stated that the contents of the course have been slightly too challenging.

\subsection{Discussion addressing our research hypotheses}
Regarding our research hypotheses established in section \ref{subsection:research_hypotheses}, we investigated that the evaluation of the questionnaires outlined that our course helped the participants to gain a basic understanding of AI in medical image analysis and to feel better able to recognise possible sources of error in the use of AI for medical image analysis. The evaluation of the questions addressing the attitude of the participants towards AI in medicine showed a belief in the hypotheses that better understanding of AI increases trust in AI. Trust in AI on the other hand encourages the willingness to use it in daily professional life.

\section{Conclusions}
In summary, our project Learn2Trust indicates that developing interactive course material that combines theoretical basics and practical application to medical image data in an accessible fashion for medical students with little to no computer science background offers great promise in building a better understanding and willingness to use deep learning techniques in clinical practice. Streamlit offers a good computational backbone for visualising theory, code and results as well as enabling users to actively engage with the data science topics. In future, it would be interesting to investigate in more detail which of our educational building blocks is most effective in teaching AI concepts and algorithms to a more general audience. 

\section*{References}

\medskip

{
\small

[1] Stanford University CS231n: Convolutional Neural Networks for Visual Recognition. {\it Retrieved May 6, 2021}, \url{http://cs231n.stanford.edu}

[2] Faes, L.\ , Wagner, S. K.\ , Fu, D. J.\ , Liu, X.\ , Korot, E.\ , Ledsam, J. R.\ , Back, T.\ , Chopra, R.\ , Pontikos, N.\ , Kern, C.\ , Moraes, G.\ , Schmid, M. K.\ , Sim, D.\ , Balaskas, K.\ , Bachmann, L. M.\ , Denniston, A. K.\ \& Keane, P. A.\ (2019) Automated deep learning design for medical image classification by health-care professionals with no coding experience: A feasibility study. {\it The Lancet Digital Health}, 1(5), e232–e242. \url{ https://doi.org/10.1016/s2589-7500(19)30108-6}

[3] Durán, J. M.\ \& Jongsma, K. R.\ (2021) Who is afraid of black box algorithms? On the epistemological and ethical basis of trust in medical AI. {\it Journal of Medical Ethics}, medethics-2020-106820. \url{https://doi.org/10.1136/medethics-2020-106820}

}




\end{document}